\documentstyle[11pt]{article}
\begin{document}
\setlength{\unitlength}{1mm}
\pagestyle{plain}
\begin{center}
{\bf Longwave Interface Instability In Two-Fluid Vibrational Flow}
\footnote{English translation from "Hydrodynamika", Perm, 1998, pp. 191 - 196}\\
\bigskip
Mikhail V. Khenner\footnote{e-mail: khenner@math.tau.ac.il} \\
{\it Computer Science Department, Tel Aviv University, Israel}\\
Dmitrii V. Lyubimov\\ 
{\it Theoretical Physics Department, Perm State
University, Russia}
\end{center}
\bigskip
\begin{center}
{\bf Abstract}
\end{center}
\begin{quotation}
\noindent 
We consider longwave mode of the interface instability in the system 
comprising of two immiscible fluid layers. The fluids fill out plane 
horizontal cavity which is subjected to horizontal harmonic vibration. 
The analysis is performed within the framework of "high frequency of 
the vibration" approximation and the averaging procedure. The 
nonlinear equation (having the form of Newton's second law) for the 
amplitude of interface deformation is obtained by means of multiple 
scales method. It is shown that (in addition to previously detected 
quasistationary periodic solutions) the equation has a class of 
quasistationary solitary solutions.  
\end{quotation}
\bigskip
\bigskip

In experimental works by Bezdeneznykh {\it et al.}~\cite{bezdeneznykh}
and by Wolf~\cite{wolf69} for a long horizontal reservoir filled
with two immiscible viscous fluids, an interesting
phenomenon was found at the interface: the horizontal vibrations lead to
the formation of a steady relief. This formation mechanism
 has a threshold nature; it is
noteworthy that such a wavy relief appears on the interface only if the
densities of the two fluids are close enough, i.e. it does not appear for the
liquid/gas interface (free surface).
The interface is absolutely unstable if the heaviest fluid
occupies the upper layer; i.e., the horizontal vibration does
not prevent the evolution of Rayleigh-Taylor instability, in contrast to the
 vertical one which under certain conditions suppresses its evolution.
A theoretical description of this phenomenon was provided by Lyubimov
\& Cherepanov~\cite{lyubimov} within the framework of a high frequency
(of the vibration) approximation and an averaging procedure; they found that a horizontal
vibration leads to a quasistationary state i.e., a state where
the mean motion is absent but the interface oscillates with a small
amplitude (of the order of magnitude of the cavity displacement)
with respect to the steady relief. 
They also obtained the general equations and boundary conditions for mean
and pulsational parts of the fluid velocities.
The theory developed in~\cite{lyubimov} made it possible to perform
the linear stability analysis for the interface.

In the approach~\cite{lyubimov}, two parameters were assumed to be
asymptotically small simultaneously:
(i) the dimensionless thickness of the viscous skin-layers
$\delta =h^{-1}\sqrt{\nu /\omega}$, $\nu$ being the kinematic
viscosity and
(ii) the dimensionless amplitude of the vibration $\epsilon=a/h$.
In this limiting case,
the possibility of description of parametric resonant effects is absent and
only the basic instability mode (Kelvin--Helmholtz, of two counter
flows) remains.

The linear stability analysis for inviscid and viscous fluids in the
case of finite $\epsilon$ and relatively low frequencies of the vibration
was carried out analytically and numerically in~\cite{klbr},~\cite{shotz}.
The transformation 
was found which reduces the linear stability problem under inviscid 
approximation to the Mathieu equation. The parametric resonant regions 
of instability associated with the intensification of 
capillary-gravity waves at the interface and the effects due to viscous
damping were examined.

In the present work, following the approach of~\cite{lyubimov}, we make the 
analytical study for longwave interface instability
in the high-frequency vibrational field.

\vspace*{5mm}
\noindent {\large \bf 1.}
Let us consider the system of two immiscible, incompressible liquids filling rectangular
cavity of length $L$ and height $h$.
In the state of rest 
the heavy liquid (of density $\rho_1$) occupies the bottom region of height $h_{1}$,
and the light liquid (of density $\rho_2$) -- the upper region of height $h_{2}$
($h=h_1+h_2$).
We choose Cartesian coordinate system in such a way that the $x,y$-axis lie in horizontal plane,
the $z$-axis is directed vertically, $z=0$ corresponds to the unperturbed interface (Fig. 1).
Let the cavity perform harmonic oscillation along the $x$-axis,
with the amplitude $a$ and frequency $\omega$.

\begin{figure}[t]
\setlength{\unitlength}{1mm}\thicklines
\begin{center}
\begin{picture}(120,50)
\put(30,53){rigid wall}
\put(30,-5){rigid wall}
\put(20,0){\line(1,0){100}}
\put(122,0){\line(1,0){2}}
\put(126,0){\line(1,0){2}}
\put(130,0){\line(1,0){2}}
\put(132,0){\line(0,1){50}}
\put(20,50){\line(1,0){100}}
\put(122,50){\line(1,0){2}}
\put(126,50){\line(1,0){2}}
\put(130,50){\line(1,0){2}}
\put(20,15){\line(1,0){100}}
\put(122,15){\line(1,0){2}}
\put(126,15){\line(1,0){2}}
\put(130,15){\line(1,0){2}}
\put(18,15){\line(-1,0){2}}
\put(14,15){\line(-1,0){2}}
\put(10,15){\line(-1,0){2}}
\put(18,0){\line(-1,0){2}}
\put(14,0){\line(-1,0){2}}
\put(10,0){\line(-1,0){2}}
\put(18,50){\line(-1,0){2}}
\put(14,50){\line(-1,0){2}}
\put(10,50){\line(-1,0){2}}
\put(8,0){\line(0,1){50}}
\put(94,0){\line(0,1){50}}
\put(60,18){interface}
\put(52,2){1}
\put(52,37){2}
\put(20,15){\vector(0,1){20}}
\put(20,15){\vector(1,0){20}}
\put(20,15){\vector(1,1){15}}
\put(16.5,33){z}
\put(38,16.5){x}
\put(35,27){y}
\put(19,12.5){o}
\put(134,15){\vector(0,1){35}}
\put(134,50){\vector(0,-1){35}}
\put(136,37){$h_{2}$}
\put(134,0){\vector(0,1){15}}
\put(134,15){\vector(0,-1){15}}
\put(136,7){$h_{1}$}
\put(-10,25){\vector(1,0){15}}
\put(5,25){\vector(-1,0){15}}
\put(-9,27){$a\cos{\omega t}$}
\end{picture}\\
\bigskip
\qquad Fig. 1. \quad Problem configuration.
\end{center}
\end{figure}

In the basic state (which is a counter flow), the interface could be considered as 
plane and horizontal. For discussion of this issue as well as the approximation
of infinite horizontal layer we refer
the readers to~\cite{klbr}.

The quasistationary perturbed state is found from the following problem:

$$
\triangle \Psi = 0, \qquad \triangle \Phi = 0,
\eqno{(1)}
$$

$$
z = -H_1: \quad \Psi = 0; \qquad z = H_2: \quad \Phi = 0,
\eqno{(2)}
$$

$$z = \xi(x):$$ 

$$
\Psi - \Phi = {(\rho - 1)(H_1 + H_2) \over {\rho H_2 + H_1}}\xi,
\eqno{(3)}
$$

$$
\rho(\Psi_z - \Psi_x \xi_x) = \Phi_z - \Phi_x \xi_x,
\eqno{(4)}
$$

$$
B\left[{\rho(H_1+H_2) \over {\rho H_2+H_1}}\Psi_z
+ {H_1+H_2 \over {\rho H_2+H_1}}\Phi_z + \Psi_z \Phi_z + \Psi_x \Phi_x \right]
- \xi + {\xi_{xx} \over {(1+\xi_x^2)^{3/2}}} = const,
\eqno{(5)}
$$

$$
\rho = \rho_1/\rho_2, \qquad B = {a^2\omega^2 \over {4}}\left({\rho_1-\rho_2 \over
{\alpha g}}\right)^{1/2}, \qquad H_1=h_1/L, \qquad H_2=h_2/L,
$$

\noindent where $\Psi$ and $\Phi$ are the streamfunctions of small 2D normal perturbations,
$\alpha$ is the coefficient of the interface tension, $\xi$ is interface deformation. Equations (1)-(5) are in dimensionless
form; the length scale is $L=[\alpha/(\rho_1-\rho_2)g]^{1/2}$ and this is also the scale
for $\Psi$ and $\Phi$. The $x,z$ differentiation is denoted by the respective subscripts.
In the case of equal heights of the layers ($H_1=H_2=H$), the equations (1)-(5)
are reduced to equations (2.6)-(2.7) in~\cite{lyubimov}.

\vspace*{5mm}
\noindent {\large \bf 2.}
Let us consider the amplitude $\xi(x)$ of interface deformation to be small,
$\xi \ll 1$. This allows to impose boundary conditions on the unperturbed
interface, $\xi=0$, being accurate up to linear terms. 
According to the main idea of multiple scales method, we introduce the
set of lengths,

$$
x_1=\epsilon x, \qquad x_2 =\epsilon^2 x, \qquad x_3 =\epsilon^3 x,...
$$

\noindent and we assume that all variables in (1)-(5) are the functions of these lengths.
Then we have the following expansions:

$$
{\partial f \over {\partial x}} = \epsilon {\partial f \over {\partial x_1}}+
\epsilon^2 {\partial f \over {\partial x_2}} +
\epsilon^3 {\partial f \over {\partial x_3}} + ..., \qquad
\Psi = \Psi_2\epsilon^2 + \Psi_4\epsilon^4 + ...,
\eqno{(6)}
$$

$$
\Phi = \Phi_2\epsilon^2 + \Phi_4\epsilon^4 + ..., \qquad
\xi = \xi_2\epsilon^2 + \xi_4\epsilon^4 + ...
$$

\bigskip
\noindent The parameter $B$, characterizing the vibration intensity, is represented like
$B = B_* +\epsilon^2 r,$ $B_*$ being the threshold instability value with respect to
longwave perturbations, 
$r$ is super(under)criticality parameter.

From (1)-(5) we get the following relations in the leading order of the expansion 
in $\epsilon$:

$$
\Psi_2 = {(\rho-1)(H_1+H_2)H_1 \over {(\rho H_2 + H_1)^2}}\xi_2
\left(1+{z \over {H_1}}\right), \qquad
\Phi_2 = {\rho(1-\rho)(H_1+H_2)H_2 \over {(\rho H_2 + H_1)^2}}\xi_2
\left(1-{z \over {H_2}}\right),
\eqno{(7)}
$$

$$
B_* = {(\rho H_2 + H_1)^3 \over {2\rho(\rho-1)(H_1+H_2)^2}}.
\eqno{(8)}
$$

\bigskip
\noindent
The equation (8), in case of equal heights of the layers, is reduced to 

$$
B_* = {H(\rho+1)^3 \over {8\rho(\rho-1)}},
$$

\noindent which is the correct instability threshold value~\cite{lyubimov}.

In the next order we have (instead of $\xi_2$ we write just $\xi$):

$$
\Psi_4 = C_1 z^3 + C_2 z^2 +
C_3 z + C_4, \qquad
\Phi_4 = C_5 z^3 + C_6 z^2 +
C_7 z + C_8,
\eqno{(9)}
$$

$$
\xi_{xx}\left(1-{H_1^3+\rho H_2^3 \over {3(\rho H_2 + H_1)}}\right) +
{2\rho(\rho-1)(H_1+H_2)^2 \over {(\rho H_2 + H_1)^3}}r \xi +
{3(\rho-1) \over {\rho H_2 + H_1}}\xi^2 = C, \qquad C = const.
\eqno{(10)}
$$

\bigskip
\noindent Here $C_1-C_8$ are known functions of $\rho,H_1,H_2,\xi(x)$.

It was shown in~\cite{lyubimov} that in thin ($H^2 < 3$) layers of equal
thickness $H$ the most dangerous are longwave perturbations
(in the sense that they appear at the smallest
possible destabilization amplitude). The same longwave perturbations
in the layers of different heights are the most dangerous if
$(H_1^3+\rho H_2^3)/(\rho H_2+H_1) < 3.$
In the following, we consider this case only. Besides, we are interested
only in the solutions to (10) which are zero at infinity, i.e. for
such solutions $C=0$.

The nonlinear equation (10) could be rewritten in the form of Newton's 2nd law:

$$
\xi_{xx} = - dU/d\xi, \qquad  U = {1 \over {2}}qr\xi^2 + {1 \over {3}}p\xi^3,
\quad q(\rho,H_1,H_2) > 0, \quad p(\rho,H_1,H_2) > 0.
\eqno{(11)}
$$

\bigskip
\noindent
The function $U(\xi)$ (potential energy) 
is presented in Fig. 2, for different values $r < 0$ $(r=-0.4,\;-0.6,\;-0.75)$.

To zero level of $U$ corresponds the deformation of the interface which
has the form of quasi-stationary soliton (Fig. 3).
It's amplitude, as follows from (11), is given by

$$
\xi_m = {3 \over {2}}{q \over {p}}|r|.
$$

\bigskip
\noindent
The soliton is stable, since the values $r<0$ correspond to undercriticality.
This gives hope to observe it in the future experiments.

It is noteworthy that besides the solutions of solitary type to eq. (11), 
in the space of parameters we find also periodic solutions, which
correspond to values $U<0$. These solutions were examined in details
in~\cite{lyubimov},~\cite{zamaraev}.


\begin{thebibliography}{200}

\bibitem{bezdeneznykh}
N.A. Bezdenezhnykh et al., 1984. Control of the fluid interface stability by
vibration, electric and magnetic fields. {\it III All-Union Seminar on
hydromechanics and heat/mass transfer in microgravity}. Chernogolovka,
p. 18-20 (in russian).

\bibitem{klbr}
M.V. Khenner, D.V. Lyubimov, T.S. Belozerova, B. Roux,
Stability of plane-parallel vibrational flow in a two-layer system,
accepted to {\it European Journal of Fluid Mechanics}. The preprint, featuring
the main results of this work could be found at {\it http://xxx.lanl.gov/}
(physics/9908007, 04.08.1999).

\bibitem{lyubimov}
D.V. Lyubimov, A.A. Cherepanov, 1987. Development of a steady relief
at the interface of fluids in a vibrational field. {\it Fluid Dynamics}\ {\bf 22},
p. 849-854.

\bibitem{shotz}
D.V.Lyubimov, M.V.Khenner, M.M.Shotz, 1998. Stability of a
fluid interface under tangential vibrations.
{\it Fluid Dynamics}\ {\bf 33}(3), p.318-323.

\bibitem{wolf69}
G.H.Wolf, 1969. The dynamic stabilization of the Rayleigh -Taylor
instability and the corresponding dynamic equilibrium. {\it Z. Physik}\ {\bf
227}(3), p. 291-300.

\bibitem{zamaraev}
A.V. Zamarayev, D.V. Lyubimov, A.A. Cherepanov, 1989.
On equilibrium interface shapes in vibrational field.
{\it UB-AS USSR, Hydrodynamics and heat-mass transfer}, p. 23-28
(in russian).

\end{thebibliography}
\end{document}